\documentclass[prd,aps,showpacs,eqsecnum,twocolumn,superscriptaddress,amsfonts]{revtex4}
\usepackage{amsmath, amssymb}
\usepackage{latexsym}
\usepackage{graphicx}
\usepackage{epsfig}
\newtheorem{Def}{Def.}
\newtheorem{Prop}{Prop.}

\newcommand{\si}{\mathcal{S}}
\newcommand{\pim}{\mathcal{P}}
\newcommand{\qum}{\mathcal{Q}}

\begin{document}

%%%%%%%%%%%%%%%%%
%%%   TITLE   %%%
%%%%%%%%%%%%%%%%%

\title{Towards a Wave-Extraction Method for Numerical Relativity:\\II.~The
quasi-Kinnersley Frame}

%%%%%%%%%%%%%%%%%%%
%%%   AUTHORS   %%%
%%%%%%%%%%%%%%%%%%%

\author{Andrea~Nerozzi}
\affiliation{Institute of Cosmology and Gravitation,
University of Portsmouth, Portsmouth PO1 2EG,  UK}
\affiliation{Center
for Relativity, Department of Physics, University of Texas
at Austin, Austin, Texas 78712-1081}
\author{Christopher~Beetle}
\affiliation{Department of Physics, University of Utah,
Salt Lake City, Utah 84112}
\affiliation{Department of Physics, Florida Atlantic University,
Boca Raton, Florida 33431}
\author{Marco~Bruni}
\affiliation{Institute of Cosmology and Gravitation,
University of Portsmouth, Portsmouth PO1 2EG,  UK}
\author{Lior~M.~Burko}
\affiliation{Department of Physics, University of Utah,
Salt Lake City, Utah 84112}
\affiliation{Department of
Physics and Astronomy, Bates College, Lewiston, Maine 04240}
\author{Denis~Pollney}
\affiliation{Max-Planck-Institut f\"ur Gravitationsphysik,
Albert-Einstein-Institut, 14476 Golm, Germany}

%%%%%%%%%%%%%%%%
%%%   DATE   %%%
%%%%%%%%%%%%%%%%

\date{July 2, 2004}

%%%%%%%%%%%%%%%%%%%%
%%%   ABSTRACT   %%%
%%%%%%%%%%%%%%%%%%%%

\begin{abstract}
The Newman--Penrose formalism may be used in numerical relativity to 
extract
coordinate-invariant information about gravitational radiation emitted in 
strong-field
dynamical scenarios. The main challenge in doing so is to identify a null 
tetrad appropriately
adapted to the simulated geometry such that Newman--Penrose quantities 
computed
relative to it have an invariant physical meaning.  
In black hole perturbation theory,
the Teukolsky formalism uses such adapted tetrads, those which differ only
perturbatively from the background Kinnersley tetrad.  At late times,
numerical simulations of astrophysical processes producing isolated black
holes ought to admit descriptions in the Teukolsky formalism.
However, adapted tetrads in this context must be identified using only
the numerically computed metric, since no background Kerr geometry
is known \textit{a priori}.  To do this, this paper introduces
the notion of a quasi-Kinnersley frame.  This frame, when space-time
is perturbatively close to Kerr, approximates the background Kinnersley 
frame.
However, it remains calculable much more generally, in space-times
non-perturbatively different from Kerr.  We give an explicit solution
for the tetrad transformation which is required in order to find this 
frame in a general space-time.
\end{abstract}

%%%%%%%%%%%%%%%%
%%%   PACS   %%%
%%%%%%%%%%%%%%%%

\pacs{
04.25.Dm, % numerical relativity
04.30.Db, % gravitational wave generation and sources
04.70.Bw, % classical black holes
95.30.Sf, % relativity and gravitation
97.60.Lf  % black holes (astrophysics)
}

%%%%%%%%%%%%%%%%%%%%%
%%%   MAKETITLE   %%%
%%%%%%%%%%%%%%%%%%%%%

\maketitle

%%%%%%%%%%%%%%%%%%%%%%%%
%%%   INTRODUCTION   %%%
%%%%%%%%%%%%%%%%%%%%%%%%

\section{Introduction}
\label{sec:introduction}

One of the main challenges currently faced by numerical relativity is that
of interpreting its results in a physically meaningful way.
That is, once a given simulation is complete, one must find ways to
quantify invariantly the physical information contained in the
gravitational field described by the numerical variables.
A new generation of experiments designed to detect and interpret
gravitational radiation has lent particular importance to one such
problem: that of extracting information about gravitational waves
far from a modeled source.  A great deal is known about gravitational
radiation in various approximation schemes, such as the standard
quadrupole formula of linearized gravity and the various approaches
(Regge--Wheeler \cite{Regge57}, Zerilli \cite{Zerilli70a}, Teukolsky 
\cite{Teukolsky73})
to black hole perturbation theory.  However, these theories are 
well-defined
only in the perturbative regime.  Each is founded on an assumed knowledge
of a specific background metric on space-time which, in a typical 
simulation of 
strongly-dynamical gravitational fields, is not known \textit{a priori}.

What is needed is a background-independent formalism which does not
rely on such \textit{a priori} structures.  Rather, one should seek
an approach based on quantities which are calculable solely from
the physical metric, and which yield information about gravitational
radiation in those cases where such radiation is unambiguously present.
Since the quantities we imagine here would be defined in terms of
the physical metric, they could in principle be calculated at any
point of any space-time.  In generic situations, however,
their interpretation in terms of gravitational radiation would be lost.

Calculations of the Newman--Penrose Weyl scalar $\Psi_4$ have 
been used in numerical studies of gravitational wave forms 
\cite{Smarr79, Anninos95g, Alcubierre01a, Baker:2001sf}.
This technique looks very promising because Weyl scalars are
first of all coordinate independent quantities. In addition, once
a suitable tetrad is found, extracting $\Psi_4$ one has immediately
the interpretation in terms of the outgoing radiation.

For such an analysis, an appropriate Newman--Penrose tetrad must be found.
This paper aims to address the problem of finding the right tetrad
to calculate $\Psi_4$.  That is, given \textit{only} the output
of a numerical simulation, we construct a particular null frame.
For space-times which truly describe perturbations of a Kerr
background, our frame approximates the Kinnersley frame of that
background.  However, the construction works somewhat more generally,
and can be applied to many numerical space-times, including
some which differ from Kerr non-perturbatively.  Specifically the tetrad
we seek belongs to one of the \textsl{transverse frames} introduced in
\cite{Beetle02}.  While three such frames exist in algebraically 
general space-times, only one can approximate the Kinnersley
frame when the space-time is a perturbation of Kerr.
Here, we show how to calculate this physically interesting 
\textsl{quasi-Kinnersley frame}. this is meant to be one
of the two steps required to have the right quantities computed
in the Teukolsky formalism, the second one being related to
fixing the scalings of the vectors constituting such frames
(see \cite{Beetle04} for further details),
in order to get the right radial fall-offs for the relevant
quantities such as $\Psi_0$ and $\Psi_4$. This second step
will be the subject of future work.
Once this construction is complete, the goal is to deploy the entire
Teukolsky formalism of black hole perturbation theory in the
weak-field radiation zones of a numerical evolution.

This paper constructs the quasi-Kinnersley frame within
the Newman--Penrose formalism.  That is, it operates by 
transforming a given, fiducial null tetrad on space-time
to one satisfying the transversality conditions.  
Because the Teukolsky formalism is built on the
Newman--Penrose approach, our results take a particularly 
clear form in this language.  However, many numerical relativity
codes do not currently incorporate the infrastructure
needed to define and transform Newman--Penrose frames on space-time.
Rather, many are based on various 3+1 decompositions of the 
Einstein equations wherein the quantities of interest describe
a spatial geometry evolving in parameter time.
This approach is meant to be alternative to the one
presented in \cite{Beetle04}, hereafter Paper I, where
the quasi-Kinnersley frame is constructed \textit{ab initio},
starting from spatial, rather than space-time, data.
Although both papers aim at the same goal, their techniques
are rather different. We present them separately to preserve clarity in
each. The issue of the quasi-Kinnersley frame is also presented
in \cite{Beetle05} where this frame is explicitly found, together with
the radiation scalar \cite{Beetle02}, for some specific cases.

The outline of this paper is as follows.  
Section \ref{sec:weyl} establishes notation and gives
general definitions, including those of both transverse 
and quasi-Kinnersley frames.
Sections \ref{sec:trans} and \ref{sec:find} 
set up and solve the problem of calculating the
three transverse frames in an algebraically general space-time.  
Section \ref{sec:genkin} shows how to select the unique quasi-Kinnersley
frame from among those three transverse frames.  
Section \ref{sec:simple} will test the construction of the
quasi-Kinnersley frame in a simple case.  
Finally, appendix \ref{app:mc} gives closed-form expressions
for the Weyl scalars and for the tetrad vectors when the fiducial
frame is the principal null frame, while appendix \ref{app:as} discusses
the existence and plurality of transverse frames in 
algebraically special space-times.

%%%%%%%%%%%%%%%%%%%%%%%%
%%%   WEYL SCALARS   %%%
%%%%%%%%%%%%%%%%%%%%%%%%

\section{Definitions}
\label{sec:weyl}

\subsection{Weyl scalars}
\label{sub:weyldef}

In vacuum space-times, curvature is entirely encoded in the Weyl tensor 
$C_{abcd}$.  The ten independent components of this tensor can be 
expressed in the five complex Weyl scalars
\begin{subequations}
\label{weyldef}
\begin{eqnarray}
\Psi_0 &=& C_{pqrs}\ell^pm^q\ell^rm^s \label{weyl0} \\
\Psi_1 &=& C_{pqrs}\ell^pn^q\ell^rm^s \label{weyl1} \\
\Psi_2 &=& C_{pqrs}\ell^pm^q\bar{m}^rn^s \label{weyl2}\\
\Psi_3 &=& C_{pqrs}\ell^pn^q\bar{m}^rn^s \label{weyl3} \\
\Psi_4 &=& C_{pqrs}\bar{m}^pn^q\bar{m}^rn^s \label{weyl4}, 
\end{eqnarray}
\end{subequations}
where $\ell^p$, $n^p$, $m^p$ and $\bar{m}^p$ comprise a null tetrad.
The first pair of vectors here are real, while the second pair are both 
complex and conjugate to one another.  The only non-vanishing inner 
products are
$\ell^p n_p=-1$ and $m^p\bar{m}_p=1$.
Relative to this non-coordinate basis, the Weyl scalars are naturally
coordinate independent, but they do depend on the particular tetrad 
choice.
The freedom in the tetrad is given by the six-dimensional Lorentz group
which, in this context, is conveniently generated by elementary
transformations of three types. For an exhaustive presentation of
these transformations we refer to Appendix \ref{sec:transform}.

Despite the complicated appearance of some of the transformation laws for 
the Weyl scalars, some combinations of them are independent of the tetrad.  
For example, two well-known scalar curvature invariants are defined by
\begin{subequations}
\label{ijdefweyl}
\begin{eqnarray}
I &=& \frac{1}{16}\left({C_{pq}}^{rs}{C_{rs}}^{pq}-
                    i{C_{pq}}^{rs}{}^*{C_{rs}}^{pq}\right) 
      \label{idefweyl} \\
J &=& \frac{1}{96}\left({C_{pq}}^{rs}{C_{rs}}^{mn}{C_{mn}}^{pq} -
      {C_{pq}}^{rs}{C_{rs}}^{mn}{}^*{C_{mn}}^{pq}\right), \nonumber \\
      \label{jdefweyl} 
\end{eqnarray}
\end{subequations}
where ${}^*{C_{pq}}^{rs} = \frac{1}{2}{\epsilon_{pq}}^{mn}{C_{mn}}^{rs}$ 
is the Hodge dual of the Weyl tensor.  By definition $I$ and $J$ do not 
depend on tetrads.  However, they can easily be expressed in terms of the 
Weyl scalars in an arbitrary tetrad:
\begin{subequations}
\label{ijdef}
\begin{eqnarray}
I &=& \Psi_4\Psi_0-4\Psi_3\Psi_1+3\Psi^2_2 \label{idef}\\
J &=& det\left|\begin{array}{ccc}
\Psi_4 & \Psi_3 & \Psi_2 \\
\Psi_3 & \Psi_2 & \Psi_1 \\
\Psi_2 & \Psi_1 & \Psi_0 \end{array}\right|.
\label{jdef}
\end{eqnarray}
\end{subequations}
For more details we refer to \cite{Chandra83,Kramer80}.

\subsection{Principal Null Directions and Additional Scalar Quantities}
\label{sec:nulldir}

Every curvature tensor picks out a family of
preferred principal null directions; principal null directions
are those preferred directions for which $\Psi_0$ or 
$\Psi_4$ are vanishing. More specifically, $\ell$ is a
principal null direction if $\Psi_0=0$ while $n$ is a 
principal null direction if $\Psi_4=0$ (see \cite{Gunnarsen95}
and \cite{dInverno71} for further details).
Since these directions are determined invariantly, 
they are natural structures to consider for the type of
tetrad construction we contemplate here.  
In this section, we review the process of identifying the
principal null directions starting from a fiducial tetrad.
This process introduces a number of quantities whose definitions will be 
important below.

The equation to be solved to find the principal directions sets $\Psi_4$ 
($\Psi_0$) to zero after an $n$ ($\ell$) null vector rotation:
\begin{equation}
{a^*}^4\Psi_0+4{a^*}^3\Psi_1+6{a^*}^2\Psi_2+4a^*\Psi_3+\Psi_4  = 0.
\label{eqn:princ}
\end{equation}
Provided we have not started in a frame where $\ell$ is already a 
principal null vector, so $\Psi_0\ne0$, we introduce the new reduced 
variable
\begin{equation}
z=\Psi_0a^*+\Psi_1,
\label{eqn:reduced}
\end{equation}
so that Eq.~(\ref{eqn:princ}) becomes the reduced equation
\begin{equation}
z^4+6Hz^2+4Gz+K  = 0.
\label{eqn:princ2}
\end{equation}
Here, $H$, $G$ and $K$ are 
\begin{subequations}
\label{hgkdef}
\begin{eqnarray}
H &=& \Psi_0\Psi_2 - \Psi_1^2 \label{hdef} \\
G &=& \Psi_0^2\Psi_3-3\Psi_0\Psi_1\Psi_2+2\Psi_1^3 \label{gdef} \\
K &=& \Psi_0^2I -3H^2. \label{kdef}
\end{eqnarray}
\end{subequations}
They can be related directly to the curvature invariants $I$ and $J$ using
\begin{subequations}
\label{ijdef2}
\begin{eqnarray}
\Psi_0^2 I &=& K+3H^2 \label{idef2} \\
\Psi_0^3 J &=& HK-H^3-G^2. \label{jdef2}
\end{eqnarray}
\end{subequations}

Unlike $I$ and $J$, the new quantities $H$, $G$ and $K$ take different 
values in different tetrads.  The solution for the principal null 
directions is then achieved introducing three additional quantities 
$\alpha$, $\beta$ and $\gamma$ defined by
\begin{subequations}
\label{alphabetagammadef}
\begin{eqnarray}
\alpha^2&=&2\Psi_0\lambda_1-4H \label{alphadef} \\
\beta^2&=&2\Psi_0\lambda_2-4H \label{betadef} \\
\gamma^2&=&2\Psi_0\lambda_3-4H, \label{gammadef}
\end{eqnarray}
\end{subequations}
where the $\lambda$ variables are the eigenvalues of a specific matrix $Q$ 
built from the Weyl scalars (see \cite{Kramer80} for further details).  
They are given by
\begin{subequations}
\label{lambdadef}
\begin{eqnarray}
\lambda_1&=& -\left(P+\frac{I}{3P}\right) \label{lambda1}\\
\lambda_2&=& -\left(e^{\frac{2\pi i}{3}}P+e^{\frac{4\pi 
i}{3}}\frac{I}{3P}\right
)\label{lambda2} \\
\lambda_3&=& -\left(e^{\frac{4\pi i}{3}}P+e^{\frac{2\pi 
i}{3}}\frac{I}{3P}\right
), \label{lambda3}
\end{eqnarray}
\end{subequations}
where 
\begin{equation}
P=\left[J+\sqrt{J^2-\left(I/3\right)^3}\right]^{\frac{1}{3}}.
\label{eqn:pdefinition}
\end{equation}
Eq.~(\ref{eqn:pdefinition}) may lead to some ambiguity.  It is easy to see 
that different choices of the branch of the cubic root permute the 
definitions for the $\lambda_i$ variables.   
The breaking of this permutation symmetry is  
essential to the definition of the quasi-Kinnersley frame \cite{Beetle04}.

In the end, we find four solutions for Eq.~(\ref{eqn:princ2}) which are
\begin{eqnarray}
z_1&=&\left(\alpha+\beta+\gamma\right)/2 \nonumber \\
z_2&=&\left(\alpha-\beta-\gamma\right)/2 \nonumber \\
z_3&=&\left(-\alpha+\beta-\gamma\right)/2 \nonumber \\
z_4&=&\left(-\alpha-\beta+\gamma\right)/2 \nonumber,
\end{eqnarray}
and the solutions of Eq.~(\ref{eqn:princ}) are easily derived from them 
using Eq.~(\ref{eqn:reduced}).

The triples of quantities $(\alpha, \beta, \gamma)$ and $(H, G, K)$ are 
both tetrad-dependent.  In fact, there is the same amount of information 
classifying a given tetrad contained in each triple.  This assertion 
follows from the relations
\begin{subequations}
\label{alH}
\begin{eqnarray}
\alpha^2+\beta^2+\gamma^2 &=& -12H \label{alH1} \\
\alpha^2\beta^2+\alpha^2\gamma^2+\beta^2\gamma^2 &=& 36H^2-4K 
\label{alH2}\\
\alpha\beta\gamma &=& 4G. \label{alH3}
\end{eqnarray}
\end{subequations}

The calculation described above could equally well be done by rotating a 
given tetrad to make $\ell$, rather than $n$, a principal null direction.  
The calculation is essentially the same, but we outline it here to 
introduce notation.  The operative equation to solve is
\begin{equation}
b^4\Psi_4+4b^3\Psi_3+6b^2\Psi_2+4b\Psi_1+\Psi_0  = 0.
\label{eqn:princb}
\end{equation}
In this case, assuming $\Psi_4\ne0$, we can introduce the reduced variable
\begin{equation}
\hat{z}=\Psi_4 b + \Psi_3,
\label{eqn:reducedb}
\end{equation}
to find the reduced equation
\begin{equation}
\hat{z}^4+6\hat{H}\hat{z}^2+4\hat{G}\hat{z}+\hat{K}  = 0,
\label{eqn:princ2b}
\end{equation}
where this time $\hat{H}$, $\hat{G}$ and $\hat{K}$ are defined as
\begin{subequations}
\label{hgkdefb}
\begin{eqnarray}
\hat{H} &=& \Psi_4\Psi_2 - \Psi_3^2 \label{hdefb} \\
\hat{G} &=& \Psi_4^2\Psi_1-3\Psi_4\Psi_3\Psi_2+2\Psi_3^3 \label{gdefb} \\
\hat{K} &=& \Psi_4^2I -3\hat{H}^2. \label{kdefb}
\end{eqnarray}
\end{subequations}
The procedure is in this case analogous to the one already presented, and 
it uses the definition of other variables $\hat{\alpha}$, $\hat{\beta}$ 
and $\hat{\gamma}$ which are given by
\begin{subequations}
\label{alphabetagammadefb}
\begin{eqnarray}
\hat{\alpha}^2&=&2\Psi_4\lambda_1-4\hat{H} \label{alphadefb} \\
\hat{\beta}^2&=&2\Psi_4\lambda_2-4\hat{H} \label{betadefb} \\
\hat{\gamma}^2&=&2\Psi_4\lambda_3-4\hat{H}. \label{gammadefb}
\end{eqnarray}
\end{subequations}
It is worth noticing that hatted variables are obtained from non-hatted 
ones by simply swapping $\Psi_0\leftrightarrow\Psi_4$ and 
$\Psi_1\leftrightarrow\Psi_3$.

\subsection{Null Tetrads and Null Frames}
\label{sec:nulltetradsnullframes}

Hereafter, we will adopt a terminology that clearly distincts
null frames and null tetrads, as follows

\begin{itemize}
\item A {\it null tetrad} is a specific set of two real null
vectors $\ell$ and $n$ and two complex conjugates null vectors
$m$ and $\bar{m}$.
\item A {\it null frame} is a class of null tetrads connected
by a spin/boost (type III) transformation.
\end{itemize}

\subsection{Transverse Frames}
\label{sec:deftransverse}

Although we are not interested in calculating principal null directions
the definitions given in section \ref{sec:nulldir} will help us provide a
rigorous definition of transverse frame for a general Petrov type I
space-time. 

Following \cite{Beetle02} we first define a transverse frame as

\begin{Def}
A \textsl{transverse frame} is a frame in which $\Psi_1=\Psi_3=0$.
\label{def:psi1psi3vanishing}
\end{Def}

We want to stress here the point that Def.~\ref{def:psi1psi3vanishing}
really identifies a frame, i.e. a class of tetrads, as it is
invariant under a spin/boost (type III) transformation.

A useful geometrical property of transverse frames is given
by the following proposition:

\begin{Prop}
A transverse frame for a Petrov type I space-time is a
frame which sees principal null directions in pairs, each pair being,
in the stereographic sphere, at the same angle $\theta$ and at
angles $\phi_1$ and $\phi_2$ such that $\phi_2-\phi_1=\pi$.
\label{prop:nullinpairs}
\end{Prop}

Let us note at this point that it is clear from
Eq.~(\ref{eqn:princ}) that it becomes a biquadratic if and only if
the frame is transverse.
Prop~\ref{prop:nullinpairs} can then be proved as follows:
let us assume that we are in transverse frame and
want to compute the principal null directions. Then 
Eq.~(\ref{eqn:princ}) becomes a biquadratic and therefore
if $\left(a^*\right)_1$ is
a solution, then $\left(a^*\right)_2=-\left(a^*\right)_1$ will be
another solution. Using stereographic
coordinates, i.e. writing the general solution for Eq.~(\ref{eqn:princ}) 
as
\begin{equation}
a^*=\cot\left(\frac{\theta}{2}\right)e^{i\phi},
\label{eqn:stereo}
\end{equation}
we see that this property corresponds to seeing the two principal
null directions at the same angle $\theta$ and at angles $\phi_1$
and $\phi_2$ such that $\phi_2-\phi_1=\pi$. 

To prove the equivalence of Def.~\ref{def:psi1psi3vanishing}
and Prop.~\ref{prop:nullinpairs} 
in the other direction let us suppose that we are in a frame
in which our parameters to get the principal null directions
have the property described in Prop.~\ref{prop:nullinpairs}, i.e.
we can write them down in the following way

\begin{equation}
\begin{array} {cc}
a^*_1=\cot\left(\frac{\theta_1}{2}\right)e^{i\phi_1} &
a^*_2=\cot\left(\frac{\theta_1}{2}\right)e^{i\phi_1+i\pi} \\
a^*_3=\cot\left(\frac{\theta_2}{2}\right)e^{i\phi_2} &
a^*_4=\cot\left(\frac{\theta_2}{2}\right)e^{i\phi_2+i\pi}.
\label{eqn:parameterstransverse}
\end{array}
\end{equation}
Using these values to build up the polynomial defined in 
Eq.~(\ref{eqn:princ})
we would end up with the term in $a^*$ and ${a^*}^3$ missing, this
corresponding to having $\Psi_1=\Psi_3=0$ in the frame we are in.

We will hereafter refer to the property 
introduced in Prop.~\ref{prop:nullinpairs}
as {\it seeing principal null
directions in conjugate pairs}, in order to distinguish it from
the normal principal null directions in pairs which define a
Petrov type D space-time.
Prop.~\ref{prop:nullinpairs} will be our starting point to define,
in the next section, the quasi-Kinnersley frame.
 
\subsection{The Quasi-Kinnersley Frame}
\label{sec:kinframe}

The Kinnersley frame \cite{Kinnersley69} is defined for a Petrov
type D space-time. Its definition states that

\begin{Def}
A \textsl{Kinnersley frame} for a type D space-time is a frame where
the two real tetrad null vectors coincide with the two repeated
principal null directions of the Weyl tensor.
\label{def:kinD}
\end{Def}

In his original article, Kinnersley
makes a second step with an additional condition 
that sets the spin coefficient $\epsilon$
to zero. This corresponds to fixing the additional degrees of freedom
coming from a spin/boost transformation, i.e. to identifying a particular
tetrad out of the Kinnersley frame. In this paper we will not consider
this second step, which deserves further study,
and focus our attention to finding a particular frame, i.e. a particular
class of null tetrads, which converges to the Kinnersley frame when
the space-time approaches a type D (see also Paper I for further
details).

In a type D space-time the following relations hold
\begin{equation} 
\begin{array} {ccc}
\si=1 & G=0 & K = 9H^2, 
\end{array}
\end{equation} 
where $\si$ is the speciality index defined in \cite{Baker00a}
\begin{equation}
\si=\frac{27J^2}{I^3}. \label{spec}
\end{equation}

We know that the Kinnersley frame has the additional property
that all the scalars are vanishing except $\Psi_2$, i.e.
it is also a canonical frame \cite{Pollney00} for Petrov type D.
We would like here to 
find that particular frame which converges
to the Kinnersley frame when $\si\rightarrow 1$. We will dub this
{\it quasi-Kinnersley frame} for a Petrov Type I space-time.
Our definition is then

\begin{Def}
A quasi-Kinnersley frame, for a Petrov Type I space-time,
is a frame which converges to the Kinnersley frame when
$\si\rightarrow 1$.
\label{def:quasikinframe}
\end{Def}

Let us consider a transverse frame as defined in
Prop.~\ref{prop:nullinpairs}, such that it sees the
principal null directions in conjugate pairs.
The difference between the angles $\phi$ of each pair of null
directions 
must remain fixed to $\pi$, even in the limit $\si\rightarrow 1$. On the
other hand, we
know that for $\si\rightarrow 1$ the two principal null directions
will eventually converge. The only way we can see, from our transverse
frame, the two parameters coinciding asymptotically, but keeping
the difference in $\phi$, is that their absolute value must tend
to zero. Hence, if asymptotically our parameters for finding the principal
null directions tend to zero, this means that our $\ell$ vector is
converging to the principal null directions, i.e. we are in a
quasi-Kinnersley frame.
 
\begin{figure}[!ht]
\begin{minipage}[t]{7.4cm}
\centering
\mbox{\epsfig{figure=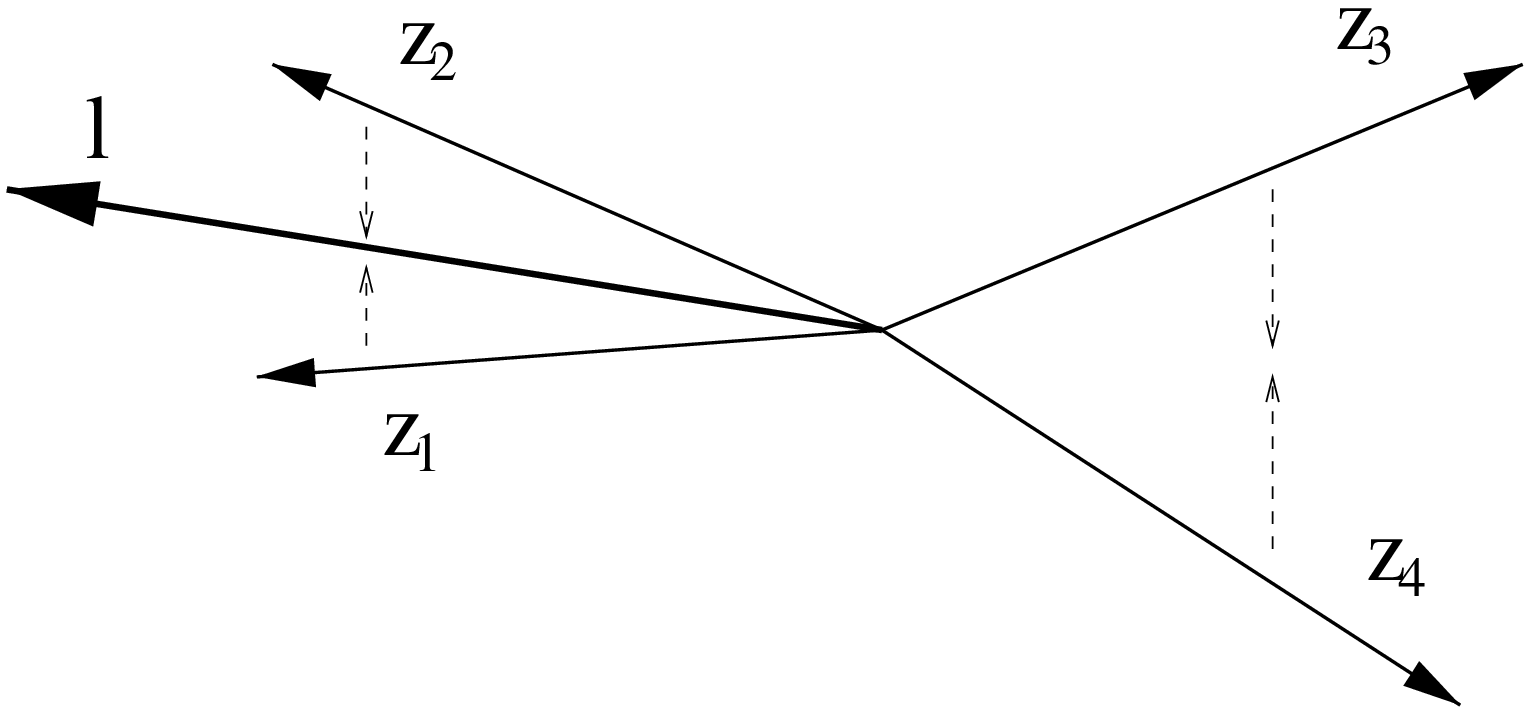,width=7.4cm, height=4.5cm}}
\caption{A transverse frame which is also a quasi-Kinnersley
frame: the $\ell$ vector of the transverse frame sees the two principal 
null directions $z_1$ and $z_2$ as conjugate pair.
As the space-time approaches a Petrov type
D one, $z_1$ and $z_2$ will converge and in particular they converge to 
$\ell$.}
\label{fig:transversekinnersley}
\end{minipage}
\hfill
\begin{minipage}[t]{7.4cm}
\centering
\mbox{\epsfig{figure=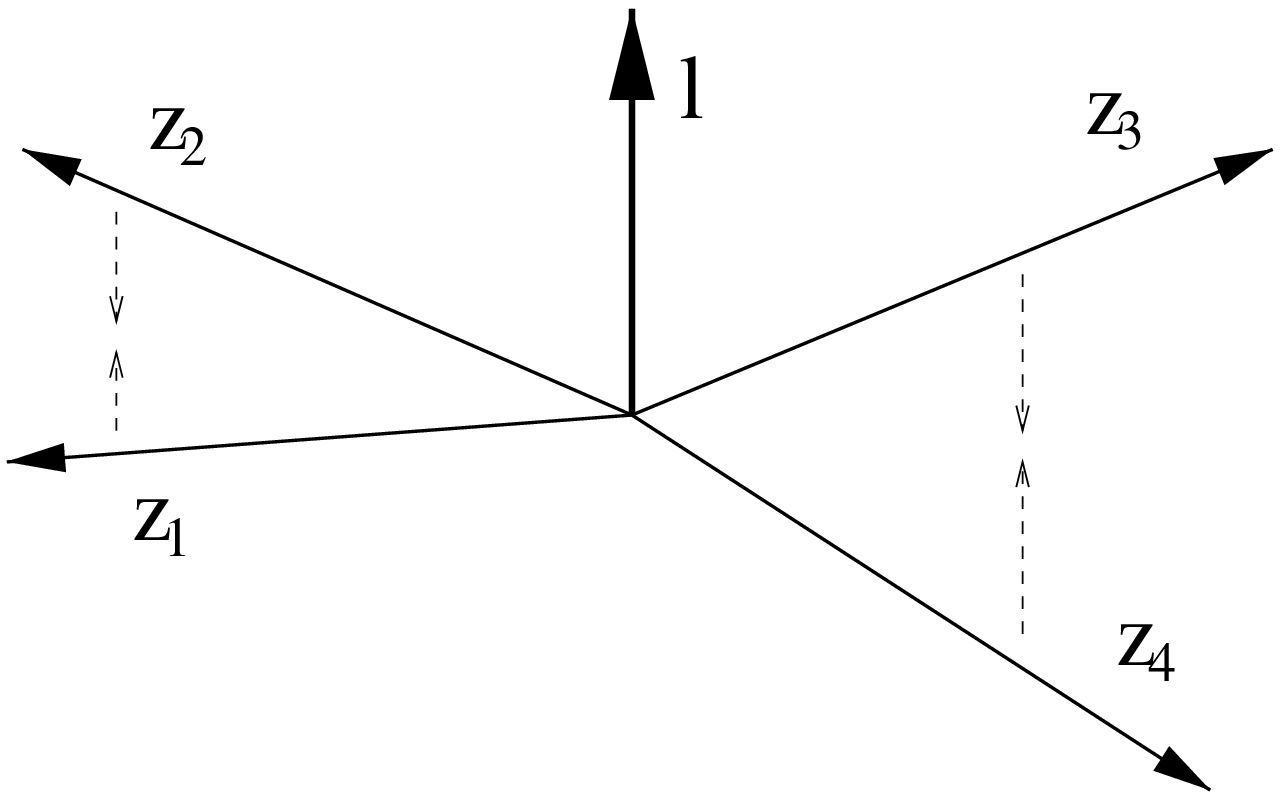,width=7.4cm, height=4.5cm}}
\caption{A transverse frame which is not a quasi-Kinnersley
frame: the $\ell$ vector of the frame sees the two principal null 
directions
$z_2$ and $z_3$ as conjugate pair. 
When $z_1$ and $z_2$ converge, they do not converge
to $\ell$.} 
\label{fig:transversenotkinnersley}
\end{minipage}
\end{figure}

Following this idea, we can conclude that a well-motivated
strategy to find a quasi-Kinnersley frame is to look for
a transverse frame.
This conclusion is however not enough. By saying that a transverse
frame sees principal null directions in conjugate pairs, we are not
specifying {\it which} directions it sees in conjugate pairs. Fig.~
(\ref{fig:transversekinnersley}) and 
Fig.~(\ref{fig:transversenotkinnersley})
explain better this concept. Let us suppose that our Petrov type I
space-time converges to a type D one, such that the principal null
directions $z_1$ and $z_2$ will converge, and the same for $z_3$
and $z_4$. In Fig.~(\ref{fig:transversekinnersley}) we have constructed
a transverse frame whose $\ell$ null vector sees 
$z_1$ and $z_2$ as conjugate pair
(which, in the graph, is indicated by putting $\ell$ in the middle
of the two principal null directions it sees in pairs); consequently
its $n$ vector will see $z_3$ and $z_4$ as conjugate pair, although this
is not shown in the figure. It turns out that this is the
quasi-Kinnersley frame as $z_1$ and $z_2$ will converge and in particular
they will converge to $\ell$. A counter example is shown in
Fig.~(\ref{fig:transversenotkinnersley}); here $l$ sees $z_2$ and
$z_3$ as conjugate pair, such that, 
when $z_1$ and $z_2$ will converge, {\em they will
not converge to $\ell$}. This is telling us that we need an additional
condition that the quasi-Kinnersley frame has to satisfy among
all the transverse frames. 

As mentioned earlier in the original Kinnersley paper the additional
condition that all the scalars are vanishing except $\Psi_2$
holds. In a general transverse frame we know from
Def.~\ref{def:psi1psi3vanishing} that $\Psi_1=\Psi_3=0$, so we want
to impose the additional condition that $\Psi_0,\Psi_4\rightarrow 0$
when $\si\rightarrow 1$ in a quasi-Kinnersley frame. By introducing
the radiation scalar $\xi=\Psi_0\Psi_4$ (notice that $\xi$ is to be
evaluated in a transverse frame; see \cite{Beetle02}) we
end up with the following proposition 

\begin{Prop}
In a Petrov type I space-time, a transverse frame where the radiation scalar
$\xi\rightarrow0$ for $\si\rightarrow1$
is a quasi-Kinnersley frame.
\label{def:kinfin}
\end{Prop} 

From the definitions given in this section and in
section \ref{sec:deftransverse} and discussion above,
it is clear that a good strategy to find a quasi-Kinnersley
frame is to search it among transverse frames, although a
transverse frame in general will not be
a quasi-Kinnersley frame.

\subsection{The linear theory}
\label{sub:linear}

Teukolsky \cite{Teukolsky73}
studied a perturbed
Kerr black hole space-time in the Newman-Penrose formalism,
choosing the Kinnersley frame for the background metric,
where for a Kerr
black hole the only non-vanishing scalar is $\Psi_2$. Having chosen
this frame, the equations governing the dynamics of all the scalars
simplify considerably, thus leading to separate evolution equations
(Teukolsky equation) for $\Psi_0$ and $\Psi_4$.
It turns out that, within the linearized
framework, i.e. considering infinitesimal transformations of the
original Kinnersley background, the
values of $\Psi_0$ and $\Psi_4$ are invariant under gauge or tetrad
transformations, 
so that they can be given pure physical
interpretation of ingoing or outgoing gravitational radiation, while
$\Psi_1$ and $\Psi_3$ can be easily set to zero, thus being related
to gauge degrees of freedom. $\Psi_2$ is instead related to the
background metric. An analogous interpretation for the scalars, not 
restricted
to linear theory, is given in \cite{Szekeres65}: here $\Psi_0$ and 
$\Psi_4$
are shown to be associated with transverse gravitational fields
(although not necessarily representing gravitational radiation), $\Psi_1$
and $\Psi_3$ to londitudinal ones, while $\Psi_2$ is related to the
Coulombian part of the gravitational field.

It is evident that if we choose the quasi-Kinnersley frame
in our numerical simulations, and we fix the particular tetrad in this 
frame
which shows the correct radial fall-offs, 
we will be able to interpret, in the linear
regime, $\Psi_4$ as the outgoing wave contribution.
Moreover, the
determination of whether we are or not in the linearized regime can
be easily achieved using the speciality index defined in Eq.~(\ref{spec})
as well described in \cite{Baker00a}.

\section{The transverse frame}
\label{sec:trans}

In the previous section we defined a transverse frame for
a Petrov type I space-time. Here we want to describe the general
problem of finding a transverse frame, as well as determining
how many transverse frames we expect.
We start from a general Petrov type I space-time
having all the five Weyl scalars non-vanishing;
we then perform an $n$ null rotation (type I) with
parameter $a$ and an $\ell$ null rotation (type II)
with parameter $b$, and set to
zero the final values of $\Psi_3$ and $\Psi_1$, ending up with a system
of two equations to be solved for parameters $a_*$ and $b$
\begin{eqnarray}
\Psi_3+3a^*\Psi_2+3{a^*}^2\Psi_1+{a^*}^3\Psi_0&+& \label{pareqn1}\\
b\left(\Psi_4+4a^*\Psi_3+6{a^*}^2\Psi_2+4{a^*}^3\Psi_1+{a^*}^4\Psi_0\right) 
&=& 0 \nonumber
\\\Psi_1+a^*\Psi_0+3b\left(\Psi_2+2a^*\Psi_1+{a^*}^2\Psi_0\right)&+& 
\label{pareqn2}\\
3b^2\left(\Psi_3+3a^*\Psi_2+3{a^*}^2\Psi_1+{a^*}^3\Psi_0\right)&+&\nonumber 
\\
b^3\left(\Psi_4+4a^*\Psi_3+6{a^*}^2\Psi_2+4{a^*}^3\Psi_1+{a^*}^4\Psi_0\right)&=&0. 
\nonumber
\end{eqnarray}

If we derive $b$ from Eq.~(\ref{pareqn1}), we get

\begin{equation}
b=-\frac{\Psi_3+3a^*\Psi_2+3{a^*}^2\Psi_1+{a^*}^3\Psi_0}{\Psi_4+4a^*\Psi_3+6{a^*}^2\Psi_2+4{a^*}^3\Psi_1+{a^*}^4\Psi_0}.
\label{exprb}
\end{equation}
This expression for $b$ is well-posed. We might be wondering
if the denominator of Eq.~(\ref{exprb}) can be vanishing. It turns out
that it cannot, as this would mean that the $n$ vector after the
$n$ null rotation (type I) with parameter $a^*$ coincides with one 
principal
null direction. From the definitions and propositions given in section
\ref{sec:deftransverse} it is clear that the $\ell$ 
and $n$ vectors of a transverse
frame do not coincide with the principal null directions for
a Petrov type I space-time. 

Substituting Eq.~(\ref{exprb}) into Eq.~(\ref{pareqn2}), 
we obtain the following sixth order
equation for the parameter $a^*$
\begin{equation}
\pim_1{a^*}^6+\pim_2{a^*}^5+\pim_3{a^*}^4+\pim_4{a^*}^3+
\pim_5{a^*}^2+\pim_6a^*+\pim_7=0,
\label{aeq}
\end{equation}
where
\begin{eqnarray} 
\pim_1 &=& -\Psi_3\Psi_0^2-2\Psi_1^3+3\Psi_2\Psi_1\Psi_0 \nonumber \\
\pim_2 &=& 
-2\Psi_3\Psi_1\Psi_0-\Psi_0^2\Psi_4+9\Psi_2^2\Psi_0-6\Psi_2\Psi_1^2
\nonumber \\
\pim_3 &=& 
-5\Psi_1\Psi_4\Psi_0-10\Psi_3\Psi_1^2+15\Psi_3\Psi_2\Psi_0\nonumber \\
\pim_4 &=& -10\Psi_4\Psi_1^2+10\Psi_3^2\Psi_0 \nonumber \\
\pim_5 &=& 5\Psi_3\Psi_0\Psi_4+10\Psi_1\Psi_3^2-15\Psi_1\Psi_2\Psi_4 
\nonumber \\
\pim_6 &=& 
2\Psi_3\Psi_1\Psi_4+\Psi_4^2\Psi_0-9\Psi_2^2\Psi_4+6\Psi_2\Psi_3^2
\nonumber \\
\pim_7 &=& \Psi_1\Psi_4^2+2\Psi_3^3-3\Psi_2\Psi_3\Psi_4. \nonumber
\end{eqnarray} 

Eq.~(\ref{aeq}) is of course very difficult  to solve analytically and we 
might 
turn to numerical methods to find solutions.

It is worth pointing
out here that we could be misled to the conclusion that we
have six transverse frames, as the equation is of sixth order. This turns
out to be wrong, due to a degeneracy of the transverse frame
if we exchange the $\ell$ and $n$ vectors: the non-vanishing
scalars would
be exchanged as follows
\begin{eqnarray}
\Psi_0 &\rightarrow& \Psi_4^* \nonumber \\ 
\Psi_2 &\rightarrow& \Psi_2 \nonumber \\
\Psi_4 &\rightarrow& \Psi_0^*, \nonumber  
\end{eqnarray}
more precisely, we would obtain a simple exchange 
$\Psi_0\leftrightarrow\Psi_4$ without complex conjugation
if we exchanged accordingly $m$ and $\bar{m}$, thus 
preserving the tetrad orientation. This is exactly the
exchange operation introduced in \cite{Beetle04}.
Although the frame we would get after such exchange
would result as a different solution of 
Eq.~(\ref{aeq}), it is actually the same from the
physical point of view, as we have just swapped the outgoing and
ingoing contribution on the scalars $\Psi_0$ and $\Psi_4$.
We will name hereafter this property as the $\ell\leftrightarrow n$
degeneracy.

We conclude then that it is possible to find three transverse
frames for a Petrov type I space-time, up to spin/boost transformations.
This result is in agreement with what was found
in \cite{Beetle02}.

Another comment to be done on Eq.~(\ref{aeq}) is that its solutions
are all we really need, as once $a$ is obtained, the parameter $b$ can
be easily derived from Eq.~(\ref{exprb}). For this reason we will no 
longer
mention the parameter $b$ from now on, and we will restrict our attention
to finding the solutions for $a$.  

\section{Finding the transverse frames}
\label{sec:find}

We will now derive the general solution for the parameter
$a$ which leads to the three transverse frames.
Our goal is to solve Eq.~(\ref{aeq}). It can be shown easily that
this equation corresponds to setting to zero the quantity $\hat{G}$
(\ref{gdefb}) after the
$n$ null rotation (type I) with parameter $a$, i.e.
\begin{equation} 
\hat{G}^a=\frac{\hat{\alpha}^a\hat{\beta}^a\hat{\gamma}^a}{4}=0, 
\label{gzero}
\end{equation}
where the index $a$ tells us that these are the quantities in the
frame which we get {\em after} the $n$ null rotation.
The equivalence of Eq.~(\ref{gzero}) with Eq.~(\ref{aeq}) is evident if,
in the substitution of Eq.~(\ref{exprb}) into Eq.~(\ref{pareqn2}), one 
does
not explicitly expand the Weyl scalars in terms of $a^*$ after the
first $n$ null rotation.

Eq.~(\ref{gzero}) expresses in a much more evident way the presence
of three transverse frames. Moreover it gives us a straightforward
way to factorize Eq.~(\ref{aeq}), as 
each of the three transverse frames 
can be defined as follows

\begin{subequations}
\label{trframes}
\begin{eqnarray}
\rm{I} &:& \hat{\alpha}^a=0 \label{trI} \\
\rm{II} &:& \hat{\beta}^a=0 \label{trII} \\
\rm{III} &:& \hat{\gamma}^a=0 \label{trIII}
\end{eqnarray}
\end{subequations}

This conclusion allows us to reduce the degree of the polynomial
originally defined in Eq.~(\ref{aeq}). Let us now focus our attention on
just one transverse frame (frame I) which verifies the condition
$\hat{\alpha}^a=0$. For the sake of simplicity, as we have defined
$\hat{\alpha}^2$ in Eq.~(\ref{alphadefb}), 
we will study the completely equivalent
condition $\left(\hat{\alpha}^a\right)^2 = 0$. 
If we write this condition in terms of the variables
in the original frame (using Eq.~(\ref{eqn:typeIrotscalars}))
we get 
\begin{equation}
\qum_1z^4+\qum_2z^3+\qum_3z^2+\qum_4z+\qum_5 = 0 \label{eqn4},
\end{equation}
where
\begin{eqnarray}
\qum_1 &=& \Psi_0\lambda_1-2H \nonumber \\
\qum_2 &=& -4G \nonumber \\
\qum_3 &=& 6\Psi_0\lambda_1H+6H^2-2K \nonumber \\
\qum_4 &=& 4G\left(H+\Psi_0\lambda_1\right) \nonumber \\
\qum_5 &=& -2KH+2G^2+\Psi_0\lambda_1K, \nonumber  
\end{eqnarray}
and $z$ is the reduced variable defined in Eq.~(\ref{eqn:reduced}).
Eq.~(\ref{eqn4}) is already a good achievement as we passed from a
sixth order equation to a fourth order one. But still this is not enough.
As mentioned previously we are actually studying the condition
$\left(\hat{\alpha}^a\right)^2 = 0$ so we want to be able to calculate
the square root of this polynomial and reduce it to a second order
equation.

Using Eq.~(\ref{alphabetagammadef}),
(\ref{hgkdef}) and (\ref{ijdef2}) it is possible to do that,
the second order polynomial being
\begin{equation}  
z^2-\left(\frac{2G}{\Psi_0\lambda_1-2H}\right)z-\left(H+\Psi_0\lambda_1\right)=0, 
\label{eqnred}
\end{equation}  
whose solutions are
\begin{equation}
z_{1,2}=\frac{G\pm 
\sqrt{G^2+\left(\Psi_0\lambda_1-2H\right)^2\left(H+\Psi_0\lambda_1\right)}}{\Psi_0\lambda_1-2H}. 
\label{eqnsol}
\end{equation}

The $\pm$ in Eq.~(\ref{eqnsol}) is related to the 
$\ell \leftrightarrow n$ degeneracy.
We can re-express Eq,~(\ref{eqnsol}) in a more elegant 
and suitable form, as a function
of the $\alpha$, $\beta$ and $\gamma$ variables. Moreover, the same 
procedure can be
applied to Eq.~(\ref{trII}) and (\ref{trIII}) in order to find
the parameter to get to the other
two transverse frames. The final result is
\begin{subequations} 
\label{finsol}
\begin{eqnarray}
\left(z\right)_{\rm{I}} &=& 
\frac{1}{2\alpha}\left[\beta\gamma\pm\sqrt{\left(\alpha^2-\beta^2\right)\left(\alpha^2-\gamma^2\right)}\right] 
\label{finsol1} \\
\left(z\right)_{\rm{II}} &=& 
\frac{1}{2\beta}\left[\alpha\gamma\pm\sqrt{\left(\beta^2-\gamma^2\right)\left(\beta^2-\alpha^2\right)}\right] 
\label{finsol2} \\
\left(z\right)_{\rm{III}} &=& 
\frac{1}{2\gamma}\left[\alpha\beta\pm\sqrt{\left(\gamma^2-\alpha^2\right)\left(\gamma^2-\beta^2\right)}\right].  
\label{finsol3}
\end{eqnarray}
\end{subequations}

The initial parameter $a^*$ can be easily found using 
Eq.~(\ref{eqn:reduced}).

\section{The quasi-Kinnersley Frame}
\label{sec:genkin}

Now that we have obtained the solutions for all the transverse
frames in a Petrov type I space-time, we wish to check if it is
possible to determine which one of them is the quasi-Kinnersley
frame we are looking for. As stated in section \ref{sec:kinframe}
this frame must satisfy the additional condition that $\xi\rightarrow 0$
when $\si\rightarrow 1$.

Our starting point are Eq.~(\ref{lambdadef}). We need to
calculate their limit when $\si\rightarrow1$. Using Eq.~
(\ref{eqn:pdefinition}) we know that $P\rightarrow J^{\frac{1}{3}}$.
In order to substitute this value into Eq.~(\ref{lambdadef})
we need to express it in function of $I$, using $I^3\rightarrow 27J^2$.
We face here again the problem of branch choosing to take the
root of a complex number; let us for the moment fix one branch
and have $J^{\frac{1}{3}}\rightarrow \left(\frac{I}{3}\right)
^{\frac{1}{2}}$. Using
this expression we get that

\begin{subequations} 
\label{eqn:asymlambdaval}
\begin{eqnarray}
\lambda_1&\rightarrow& -2\sqrt{I/3} \label{asymlambda1} \\
\lambda_2&\rightarrow& \sqrt{I/3} \label{asymlambda2} \\
\lambda_3&\rightarrow& \sqrt{I/3}. \label{asymlambda3}
\end{eqnarray}
\end{subequations} 
Eq.~(\ref{eqn:asymlambdaval}) help us remove the 
ambiguity of choosing the right
branches. No matter what branches we choose in taking roots of complex
numbers, we will end up having three $\lambda$ variables, one of which
will have a greater absolute value, precisely twice as much than
the other two, in zones of the space-time close to type D. Once
identified that particular $\lambda$ variable, we will name it
$\lambda_1$. The remaining freedom in naming $\lambda_2$ and
$\lambda_3$ is not relevant to identify the quasi-Kinnersley frame.

Using now the properties of a transverse frame
given in Def.~\ref{def:psi1psi3vanishing}
and Eq.~
(\ref{alphabetagammadef}) and (\ref{trframes}),
it can be shown that the values of $\Psi_2$
in the three transverse frames are given by
\begin{subequations}
\label{eqn:psi2transverse}
\begin{eqnarray}
\left(\Psi_2\right)_{\rm{I}} &=& \lambda_1/2 \label{eqn:psi2transverseI} 
\\
\left(\Psi_2\right)_{\rm{II}} &=& \lambda_2/2  
\label{eqn:psi2transverseII}\\
\left(\Psi_2\right)_{\rm{III}} &=& \lambda_3/2. 
\label{eqn:psi2transverseIII}
\end{eqnarray}
\end{subequations}
Moreover, using Eq.~(\ref{idef}), it is possible to
show that the radiation scalar $\xi$
has the following value in the three transverse frames

\begin{subequations}
\label{eqn:xitransverse}
\begin{eqnarray}
\left(\xi\right)_{\rm{I}} &=& \left(\lambda_2-\lambda_3\right)^2/4
\label{eqn:xitransverseI} \\
\left(\xi\right)_{\rm{II}} &=& \left(\lambda_1-\lambda_3\right)^2/4
\label{eqn:xitransverseII} \\
\left(\xi\right)_{\rm{III}} &=& \left(\lambda_1-\lambda_2\right)^2/4.
\label{eqn:xitransverseIII} 
\end{eqnarray}
\end{subequations}

Hence, using Eq.~(\ref{eqn:asymlambdaval}), we conclude that the
asymptotic values for $\xi$ in the
three transverse frames when $\si\rightarrow1$ are

\begin{subequations}
\label{eqn:asymptoticxi}
\begin{eqnarray}
\left(\xi\right)_{\rm{I}}&\rightarrow& 0 \label{eqn:asymI} \\
\left(\xi\right)_{\rm{II}}&\rightarrow& 3I/4 \label{eqn:asymII} \\
\left(\xi\right)_{\rm{III}}&\rightarrow& 3I/4, \label{eqn:asymIII}
\end{eqnarray}
\end{subequations}
this leads to our conclusion that
{\em transverse frame I is the quasi-Kinnersley
frame}, as it is the only one that matches all the
criteria given in Prop.~\ref{def:kinfin}.

%%% Begin new stuff -- CB

It is worthwhile at this point to compare the definition of the quasi-Kinnersley 
frame advanced in this paper with that contained in the companion paper \cite{Beetle04}.  
In particular, the present definition operates very simply by identifying that 
eigenvalue of the Weyl tensor with the largest modulus.  
The companion paper uses a somewhat more general definition, 
deriving a non-perturbative formula for the relevant eigenvalue which holds 
throughout the disk $|\si - 1| < 1$, and then identifying the quasi-Kinnersley 
frame as the eigenvector with that particular eigenvalue.  
In the limit $\si \to 1$, both of these definitions are entirely equivalent.  
However, it is initially not at all clear to what extent they remain equivalent 
when $\si$ differs from unity by a finite amount.  
That is, although the eigenvalues themselves are degenerate only at the 
critical points where $\si = 0$ or $\si = 1$, this guarantees nothing about 
their moduli.  There could be points within the region $|\si-1|<1$ where 
two eigenvalues differ only by a phase.  
Thus, we are led to ask what the largest neighborhood of unity in the 
$\si$-plane is in which the quasi-Kinnersley frame, 
as defined in the companion paper, is actually associated with the 
eigenvalue of largest modulus.  
The answer is quite unexpected: it is the entire disk $|\si - 1| < 1$.

\begin{figure}[!ht]
\begin{minipage}[t]{7.4cm}
\centering
\mbox{\epsfig{figure=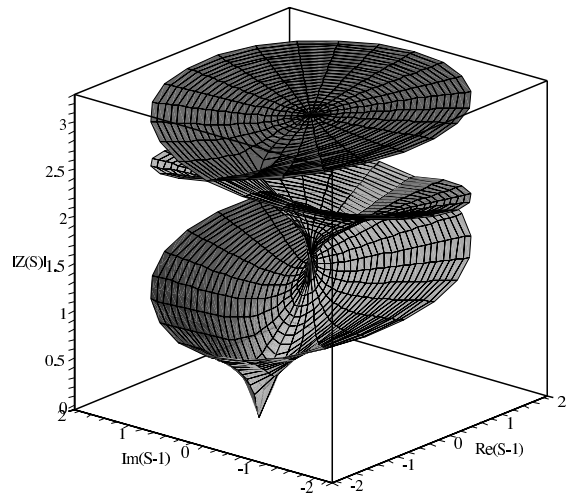,width=7.4cm}}
\caption{A representation of a function giving the 
three eigenvalues of the Weyl tensor as a function 
of $\si$ in the region $|\si-1| < 2$.  
The front lateral axis is the real part of $\si-1$, 
while the other lateral axis is its imaginary part.  
The vertical axis is the modulus of the eigenvalue, 
and the function itself is clearly triple-valued at most points.  
The figure demonstrates explicitly that the moduli 
of the eigenvalues do not equal one another except 
on the branch lines of the underlying complex function, 
where of course the eigenvalues themselves are equal.}
\label{fig:leaves}
\end{minipage}
\end{figure}

Combining Eqs.~(\ref{lambdadef}) and (\ref{eqn:pdefinition}), 
one can identify the three possible eigenvalues of the Weyl tensor 
with the three branches of a simple function of $\si$ 
(times a pre-factor involving $I$, $J$ and $\sqrt{S}$ which is 
the same for all three branches).  This is done explicitly 
in the companion paper.  At $\si = 1$, the quasi-Kinnersley frame 
is associated with the branch of largest modulus.  
Moreover, using these explicit formulae, one can plot the moduli 
of this branch alongside those of the other two in a finite 
neighborhood of unity.  
This is done in Fig.~(\ref{fig:leaves}) throughout the region $|\si - 1| < 2$.  
The topmost sheet of this surface is clearly associated 
with the quasi-Kinnersley frame at $\si=1$, 
the center of the polar coordinates used to generate the figure.  
Notably, this sheet does not intersect the other sheets, 
which give the moduli of the other two eigenvalues, 
except where $\si$ is real and non-positive.  
Thus, within the region $|\si-1| < 1$ of primary interest, 
the eigenvalue of largest modulus is always associated with 
the quasi-Kinnersley frame, as defined in the companion paper.  
Since outside of this region one encounters subtleties in the 
branch structure of this complex function which make even the 
definition of the companion paper somewhat problematic, 
we can conclude that the two definitions advanced in these papers are 
effectively equivalent.  This observation will simplify considerably 
the practical problem of identifying the quasi-Kinnersley frame.  
One need only find the largest eigenvalue of the Weyl tensor.

%%% End new stuff -- CB

\section{A simple case}
\label{sec:simple}

Let us suppose that we are already in a transverse frame and
we want to get the parameters that take us to the other two
frames. In order to simplify the calculations, let us also
fix the particular tetrad in the transverse frame for which
$\Psi_0=\Psi_4$.

Eq.~
(\ref{aeq}) simplifies enormously if we set $\Psi_1=\Psi_3=0$ and
$\Psi_0=\Psi_4$
in our
initial tetrad, and becomes
\begin{equation}
{a^*}^5-a^*=0,
\label{aeqsimpl}
\end{equation}
here, the solution $a^*=0$ indicates that we are already in a
transverse tetrad, while the corresponding tetrad which we would
get by the $\ell\leftrightarrow n$ degeneracy cannot be obtained with
a type I rotation (equivalently it could be obtained
using a parameter $a^*=\infty$), this explaining the one
order lowering of the polynomial.

The other relevant solutions are

\begin{equation}
a^* = 1, i, -1, -i. \label{eqn:simplesolresult}
\end{equation}

Such a solution allows us to derive 
another simple geometrical explanation
to the presence of three transverse frames for a Petrov type
I space-time, more directly linked to what an appropriately
chosen observer would measure.
Once we have the solution for $a^*$, using Eq.~(\ref{exprb}) we can
find the corresponding values for the $b$ parameter related
to the $\ell$ null rotation (type II), the result being

\begin{equation}
b = -1/2, i/2, 1/2, -i/2. \label{eqn:simplesolresultb}
\end{equation}
Now let us suppose that the tetrad we define in the first
transverse frame is built from a time-like vector $u$ and
three space-like vectors $e_1$, $e_2$ and $e_3$ in the usual way

\begin{subequations}
\label{eqn:tetradI}
\begin{eqnarray}
\ell^{p}_{\rm{I}} &=& \frac{1}{\sqrt{2}}\left(u^{p}+e_3^{p}\right) \\
n^{p}_{\rm{I}} &=& \frac{1}{\sqrt{2}}\left(u^{p}-e_3^{p}\right)\\
m^{p}_{\rm{I}} &=& \frac{1}{\sqrt{2}}\left(e_1^{p}+ie_2^{p}\right).
\end{eqnarray}
\end{subequations}
If we use the parameters $a^*=1$ and $b=-1/2$ to get to the second
transverse frame, we obtain the following expression for the new
tetrad vectors

\begin{subequations}
\label{eqn:tetradII}
\begin{eqnarray}
\ell^{p}_{\rm{II}} &=& \frac{1}{\sqrt{2}}\left(u^{p}-e_1^{p}\right) \\
n^{p}_{\rm{II}} &=& \frac{1}{\sqrt{2}}\left(u^{p}+e_1^{p}\right) \\
m^{p}_{\rm{II}} &=& \frac{1}{\sqrt{2}}\left(e_2^{p}+ie_3^{p}\right).
\end{eqnarray}
\end{subequations}

Where we have also used a type III rotation to re-adjust the
normalization constants. Analogously, using $a^*=i$ and $b=i/2$ we
can get to the third transverse frame, whose tetrad vectors are

\begin{subequations}
\label{eqn:tetradIII}
\begin{eqnarray}
\ell^{p}_{\rm{III}} &=& \frac{1}{\sqrt{2}}\left(u^{p}+e_2^{p}\right) \\
n^{p}_{\rm{III}} &=& \frac{1}{\sqrt{2}}\left(u^{p}-e_2^{p}\right) \\
m^{p}_{\rm{III}} &=& \frac{1}{\sqrt{2}}\left(e_1^{p}-ie_3^{p}\right).
\end{eqnarray}
\end{subequations}

Eq.~(\ref{eqn:tetradI}), (\ref{eqn:tetradII}) and (\ref{eqn:tetradIII})
show that the presence of three transverse frames corresponds
to the freedom an observer has in choosing one of the three
space-like vectors
in order to construct the two real null vectors $\ell$ and $n$. The
remaining two space-like vectors are then used to construct the
complex null vector $m$.

Following Szekeres's  gravitational compass~\cite{Szekeres65} approach, 
the
electric Weyl tensor represents the only direct curvature
contribution to the Jacobi (or in particular the geodesic deviation)
equation, and for any Petrov type I field and any transverse frame can be
expressed as \cite{Kramer80}

\begin{eqnarray}
E^{pq}&=&\mathrm{Re}(\Psi_2) e^{pq}_{\mathrm{C}}
-\frac{1}{2}\mathrm{Re}(\Psi_0+\Psi_4)e^{pq}_{\mathrm{T}+}\label{bella}
\\ &+&\frac{1}{2}
\mathrm{Im}(\Psi_0-\Psi_4)e^{pq}_{\mathrm{T}\times}, \nonumber
\end{eqnarray}
where in a frame as (\ref{eqn:tetradI})
\begin{eqnarray}
e^{pq}_{\rm{C}}&=& e^p_1 e^q_1 +e^p_2 e^q_2 -2e^p_3 e^q_3 \nonumber \\
e^{pq}_{\mathrm{T}\times} &=& e^p_1 e^q_2+e^p_2e^q_1 \nonumber \\ 
e^{pq}_{\mathrm{T}+}&=& e^p_1 e^q_1 -e^p_2 e^q_2,  \nonumber
\end{eqnarray}
respectively represent a
Coulombian and two transverse basis tensors.
  It is actually this expression for $E^{pq}$ than justify in general (and
not just in a perturbative context) the ``transverse frame" terminology: 
for
a generic tetrad with $\Psi_1\neq0$ or $\Psi_3\ne0$ there would also be
longitudinal contributions to (\ref{bella}) \cite{Szekeres65}.
For type D space-times, observers using a canonical null tetrad where only
$\Psi_2\neq 0$ (and associated orthonormal one) don't measure any 
transverse
contribution.
On the other hand, in a type I space-time any observer associated with a
transverse frame would measure transverse contributions stresses to 
his/her
gravitational compass, even when no gravitational radiation is present, as
it is clear for example from an analysis of the  Kasner\cite{Cherubini04} 
and
stationary axi-symmetric rotating neutron stars space-times 
\cite{Berti04}.
In these cases, however, the observer would unambiguously exclude the
presence of gravitational radiation by observing a zero super-energy flux
(see e.g. \cite{Belinski01}).

%%%%%%%%%%%%%%%%%%%%%%
%%%   CONCLUSION   %%%
%%%%%%%%%%%%%%%%%%%%%%

\section{Conclusions}
\label{sec:concl}

In this paper we have illustrated a method to explicitly 
construct this quasi-Kinnersley frame
within the Newman-Penrose formalism \cite{Newman62}.
First  we have provided the definition of the
quasi-Kinnersley frame for a general Petrov Type I space-time. This 
definition
allowed us to write down the basic equations that this particular
frame has to satisfy, and, eventually, to solve them. 
Using this solution it is possible to
rotate our arbitrary initial null tetrad to the quasi-Kinnersley frame. 
In this way we have completely fixed the four degrees of freedom coming
from $n$ (type I) and $\ell$ (type II) vector rotations, remaining with 
the two degrees
of freedom coming from spin/boost (type III) transformations, which
deserve further study.  Finally, in the appendices, we highlighted
further details on finding the transverse frames in the general case
and for algebraically special space-times.

While using the Newman-Penrose formalism \cite{Newman62} 
to construct the quasi-Kinnersley frame is certainly well suited
for codes using a characteristic formulation \cite{Nerozzi04b},
most numerical relativity is formulated using the 3+1
decomposition of Einstein equations. In this context it is 
therefore important to construct the quasi-Kinnersley frame
directly from the spatial geometry.
This approach to the construction of the quasi-Kinnersley 
frame is complementary to the one presented here,
and is presented in Paper I. Both approaches identify a quasi-Kinnersley
frame as one of the three transverse frames present in a Petrov
Type I space-time. The problem of understanding which transverse
frame is the quasi-Kinnersley frame is faced in both approaches and
different solutions are presented. In section \ref{sec:genkin}
we have shown that these solutions are completely equivalent not
only in a perturbative regime, but in the entire disk $\left|\si-1\right|<1$.

%%%%%%%%%%%%%%%%%%%%%%%%%%%%
%%%   ACKNOWLEDGEMENTS   %%%
%%%%%%%%%%%%%%%%%%%%%%%%%%%%

\acknowledgments
                                                                                
We are indebted to Christian Cherubini, Frank Herrmann,
Richard Price, Carlos Sopuerta,
Virginia Re and Frances White
for helpful discussions. This work
was supported by the EU Network Programme (Research Training Network
contract HPRN-CT-2000-00137),
by the National Science Foundation
through
grants No.~PHY-0244605 to the University of Utah and
No.~PHY-0400588 to Florida Atlantic University, and
by the NASA grant NNG04GL37G to the University of Texas
at Austin.

%%%%%%%%%%%%%%%%%%%%%%
%%%   APPENDIX     %%%
%%%%%%%%%%%%%%%%%%%%%%

\begin{appendix}

\section{Tetrad Transformations}
\label{sec:transform}

The six parameters of a Lorentz transformation acting on a null tetrad are
conveniently expressed in three complex parameters.
These parameters yield frame rotations of three types:

\begin{itemize}

\item {\bf $n$ vector null rotations (type I)}\\
leave $\ell$ unchanged, while the other vectors are
transformed as follows
\begin{equation}
\begin{array} {cc}
\ell\rightarrow \ell & n\rightarrow n+a^*m+a\bar{m}+aa^*\ell \\
m\rightarrow m+a\ell & \bar{m}\rightarrow \bar{m}+a^* \ell
\end{array}
\end{equation}
where $a$ is a complex parameter and $a^*$ is its complex conjugate. The 
effect
of this transformation on the Weyl scalars is
\begin{subequations}
\label{eqn:typeIrotscalars}
\begin{eqnarray}
\Psi_0 &\rightarrow& \Psi_0 \label{eqn:Ipsi0} \\
\Psi_1 &\rightarrow& \Psi_1+a^*\Psi_0 \label{eqn:Ipsi1}\\
\Psi_2 &\rightarrow& \Psi_2+2a^*\Psi_1+{a^*}^2\Psi_0 \label{eqn:Ipsi2}\\
\Psi_3 &\rightarrow& \Psi_3+3a^*\Psi_2 +3{a^*}^2\Psi_1+
                      {a^*}^3\Psi_0\label{eqn:Ipsi3}\\
\Psi_4 &\rightarrow& \Psi_4+4a^*\Psi_3+6{a^*}^2\Psi_2
       + 4{a^*}^3\Psi_1+{a^*}^4\Psi_0 \nonumber \\ \label{eqn:Ipsi4}
\end{eqnarray}
\end{subequations}

\item {\bf $\ell$ vector null rotations (type II)}\\
rotations change the tetrad vectors in the following way
\begin{equation}
\begin{array} {cc}
\ell\rightarrow \ell+b^*m+b\bar{m}+bb^*n & n\rightarrow n \\
m\rightarrow m+bn & \bar{m}\rightarrow \bar{m}+b^* n
\end{array}
\end{equation}
where $b$ is a second complex scalar quantity.  The Weyl scalars transform 
as
\begin{subequations}
\label{eqn:typeIIrotscalars}
\begin{eqnarray}
\Psi_0 &\rightarrow& \Psi_0+4b\Psi_1+6b^2\Psi_2
       +4b^3\Psi_3+b^4\Psi_4 \label{eqn:IIpsi0} \\
\Psi_1 &\rightarrow& \Psi_1+3b\Psi_2+3b^2\Psi_3+b^3\Psi_4  
\label{eqn:IIpsi1}\\
\Psi_2 &\rightarrow& \Psi_2+2b\Psi_3+b^2\Psi_4  \label{eqn:IIpsi2}\\
\Psi_3 &\rightarrow& \Psi_3+b\Psi_4  \label{eqn:IIpsi3}\\
\Psi_4 &\rightarrow& \Psi_4  \label{eqn:IIpsi4}
\end{eqnarray}
\end{subequations}

\item {\bf Spin/boost transformations (type III)}\\
rescale the vectors $\ell$ and $n$, and rotate $m$ and $\bar{m}$ in their 
complex plane:
\begin{equation}
\begin{array} {cc}
\ell\rightarrow A^{-1}\ell & n\rightarrow An \\
m\rightarrow e^{i\theta}m & \bar{m}\rightarrow e^{-i\theta}\bar{m}
\end{array}
\end{equation}
where $A$ and $\theta$ are two real scalars. Weyl scalars are modified
according to
\begin{subequations}
\label{eqn:typeIIIrotscalars}
\begin{eqnarray}
\Psi_0 &\rightarrow& A^{-2}e^{2i\theta}\Psi_0 \label{eqn:IIIpsi0} \\
\Psi_1 &\rightarrow& A^{-1}e^{i\theta} \Psi_1 \label{eqn:IIIpsi1}\\
\Psi_2 &\rightarrow& \Psi_2 \label{eqn:IIIpsi2}\\
\Psi_3 &\rightarrow& Ae^{-i\theta}\Psi_3 \label{eqn:IIIpsi3}\\
\Psi_4 &\rightarrow& A^2e^{-2i\theta}\Psi_4 \label{eqn:IIIpsi4}
\end{eqnarray}
\end{subequations}
\end{itemize}

\section{More comments on finding the transverse frames}
\label{app:mc}

As pointed out in Section \ref{sec:trans}, the six transverse frames
initially found are
$\ell\leftrightarrow n$ degenerate, so that only three independent
equivalency classes of transverse frames remain. In this Appendix we look
in greater detail into the properties of the frames under an exchange 
operation $\ell\leftrightarrow n$. To facilitate the discussion, we assume
here without loss of generality that our algebraically general space-time
is written in a principal null frame, for which $\Psi_0=0$ and
$\Psi_4=0$, which can always be done \cite{Chandra83}. This situation is 
of
little interest for numerical relativity applications, and is undertaken
in this Appendix for illustrating some of the mathematical properties of
transverse frames. In this Appendix we recapitulate the construction of
the transverse frames under the assumption that the initial frame is
the principal null one. This assumption allows us to write explicitly in
closed form the real null vectors of the transverse frame. Also, we
consider the properties of the transverse frame under the exchange
operation $\ell\leftrightarrow n$.

\subsection{Finding the transverse frames}

We assume an algebraically general space-time in the principal null
frame. We then perform two successive null rotations. The first is a class 
I
rotation (which keeps $\ell$ fixed) with parameter $a$, followed by a 
class
II rotation (which keeps $n$ fixed) with parameter $b$. (See Appendix
\ref{sec:transform} for details.) In what follows we denote the Weyl
scalars of the principal null frame by $\Psi_i$ ($i=0..4$),
$\Psi_i'$ are the Weyl scalars in the frame obtained after the first null
rotation, and $\Psi_i''$ in the frame obtained after the second null
rotation. By Def.~\ref{def:psi1psi3vanishing}, we are looking for
rotations such that both $\Psi_1''$ and $\Psi_3''$ are zero
simultaneously. 

We next use Eq.~(\ref{exprb}) for the particular case $\Psi_0=0=\Psi_4$ (for the 
principal null frame), which simplifies to 
\begin{equation}
b=-\frac{1}{2a^*}\frac{\Psi_3+3a^*\Psi_2+3{a^*}^2\Psi_1}
{2\Psi_3+3a^*\Psi_2+2{a^*}^2\Psi_1}\, ,
\end{equation}
to make $\Psi_3''=0$. Demanding next that $\Psi_1''$ too is zero, 
Eq.~(\ref{aeq}) simplifies to 
\begin{eqnarray}
\Psi_1^3{a^*}^6&+&3\Psi_1^2\Psi_2{a^*}^5+5\Psi_1^2\Psi_3{a^*}^4-
5\Psi_1\Psi_3^2{a^*}^2\nonumber \\
&-&3\Psi_2\Psi_3^2a^*-\Psi_3^3=0\, .
\label{soe}
\end{eqnarray}
The polynomial on the left hand side of Eq.~(\ref{soe}) can be easily
factored as 
\begin{equation}
(\Psi_1{a^*}^2-\Psi_3)^2(\Psi_1{a^*}^2+x_1)^2(\Psi_1{a^*}^2+x_2)^2=0\,
,
\end{equation}
where
\begin{equation}
x_1=\Psi_3+\frac{a^*}{2}
\left(3\Psi_2-\sqrt{9\Psi_2^2-16\Psi_1\Psi_3}\right)
\end{equation}
and 
\begin{equation}
x_2=\Psi_3+\frac{a^*}{2}\left(3\Psi_2+\sqrt{9\Psi_2^2-16\Psi_1\Psi_3}
\right)\, .
\end{equation}

As pointed out in Section \ref{sec:trans}, we can thus do six different
null rotations to transverse frames. For simplicity, let us first do the
rotations for which $\Psi_1{a^*}^2-\Psi_3=0$. Specifically, we can do
rotations with
$a^*=\pm\sqrt{\Psi_3/\Psi_1}$ and $b=\mp\sqrt{\Psi_1/(4\Psi_3)}$.
In the transverse frames we find that 
\begin{subequations}
\begin{eqnarray}
\Psi_0''&=&\frac{1}{8}\Psi_1\left(3\frac{\Psi_2}{\Psi_3}\mp
4\sqrt{\frac{\Psi_1}{\Psi_3}}\right)\\
\Psi_2''&=&-\frac{1}{2}\Psi_2\\
\Psi_4''&=&6\frac{\Psi_2\Psi_3}{\Psi_1}\pm
8\Psi_3\sqrt{\frac{\Psi_3}{\Psi_1}}\, .
\end{eqnarray}
\end{subequations}
For either choice of sign we find that the product $\Psi_0''\Psi_4''$ is
the same. Specifically, 
$\Psi_0''\Psi_4''=\frac{9}{4}\Psi_2^2-4\Psi_1\Psi_3$. Below, we show how
to find the remaining two transverse frames.

The remaining two vectors of the null tetrad, namely the complex null
vectors
$m$ and $\bar m$ can be easily found up to a rotation in the $m{\bar m}$
plane by solving the following 7 equations for the 8 unknown components
of
the two vectors. These equations are the conditions that the frame is
null, in addition to the normalization condition. Specifically,
$m\cdot m= {\bar m}\cdot{\bar m}=0$, $\ell\cdot m=\ell\cdot{\bar
m}=n\cdot m=n\cdot{\bar m}=0$, $m\cdot{\bar m}=1$. The indeterminate     
rotation parameter in the $m{\bar m}$ plane does not influence the two
real null vectors $\ell,n$, and affects the Weyl scalars only by a phase. 
In
particular, $\Psi_2$ and the product $\Psi_0\Psi_4$ (and also the product
$\Psi_1\Psi_3$) are invariant under spatial rotations in the $m{\bar m}$
plane (class III rotations).

\subsection{The $\ell\leftrightarrow n$ degeneracy}

In the preceding discussion we found that by setting 
$\Psi_1{a^*}^2-\Psi_3=0$ we find two transverse frames.
Next, we show that the two choices of signs correspond to the degeneracy
of $\ell\leftrightarrow n$ (up to a scale factor). 
Let us attach a subscript 1 to the choice of the sign $+$ in $a^*$, and a
subscript 2 to the choice of $-$. Doing the two null rotations, the
new real null vectors $\ell''$ and $n''$ satisfy
\begin{equation}
\ell_{1,2}''=\frac{1}{4}\ell\mp\frac{1}{4}
\frac{{\Psi_1^*}^{1/4}}{{\Psi_3^*}^{1/4}}m
\mp\frac{1}{4}\frac{{\Psi_1}^{1/4}}{{\Psi_3}^{1/4}}{\bar m}+\frac{1}{4}
\frac{{(\Psi_1\Psi_1^*)}^{1/4}}{{(\Psi_3\Psi_3^*)}^{1/4}}n
\end{equation} 
and
\begin{equation}
n''_{1,2}=n\pm\frac{{\Psi_3}^{1/4}}{{\Psi_1}^{1/4}}m\pm
\frac{{\Psi_3^*}^{1/4}}{{\Psi_1^*}^{1/4}}{\bar m}+
\frac{{(\Psi_3\Psi_3^*)}^{1/4}}{{(\Psi_1\Psi_1^*)}^{1/4}}\ell\, .
\end{equation} 

Then, we find that $n_1''=K_1\ell_2''$ and
$\ell_1''=K_1^{-1}n_2''$, where the scale factor $K_1=\frac{1}{4}
{(\Psi_1\Psi_1^*)}^{1/4}/{(\Psi_3\Psi_3^*)}^{1/4}$. That is, we
find that by choosing different signs for $a^*$ we arrive at the {\em
same} transverse null frame: we only change the roles of $\ell$ and $n$. 
Also,
the product $\Psi_0''\Psi_4''$ (a radiation scalar) is invariant under
this change of sign, although $\Psi_0''$ and $\Psi_4''$ are separately
not. 

\subsection{Finding the remaining two transverse frames}

To find the remaining transverse null frames, for simplicity let us do
null rotations on the frame we already found, instead of going back to the
principal null frame. (One could also do null rotations on the principal
null frame, using ${a^*}^2=-x_1/\Psi_1$ or ${a^*}^2=-x_2/\Psi_1$ with the
corresponding values for $b$. It is simpler, however, to find the
remaining transverse null frame from the one we already found.)
Specifically, let us assume that we are already in a
transverse null frame, which will henceforth be denoted by unprimed
quantities. Next, we do a
class I null rotation with (a new) parameter $a$ and a class II null
rotation with (a new) parameter $b$. The composition of these two null
rotation should preserve the transversality of the frame, i.e., we demand
that both $\Psi_1''=0$ and $\Psi_3''=0$ simultaneously. Substituting 
$\Psi_1=0=\Psi_3$ in Eq.~(\ref{exprb}), we find that the parameter 
\begin{equation}
b=-a^*\frac{3\Psi_2+{a^*}^2\Psi_0}{\Psi_4+6{a^*}^2\Psi_2+{a^*}^4\Psi_0}
\end{equation}
makes $\Psi_3''=0$. We also find that Eq.~(\ref{pareqn2}) reduces under this 
situation to 
\begin{equation}
\Psi_1''=\frac{a^*\left(9\Psi_2^2-\Psi_0\Psi_4\right)\left(\Psi_0{a^*}^4
-\Psi_4\right)}{(\Psi_4+6{a^*}^2\Psi_2+{a^*}^4\Psi_0)^2}\, .
\end{equation}

The requirement that $\Psi_1''=0$ yields 
\begin{equation}
{a^*}^4=\frac{\Psi_4}{\Psi_0}\, .
\end{equation}
(The case $9\Psi_2^2=\Psi_0\Psi_4$ which also nullifies $\Psi_1''$
degenerates to Petrov type-D space-time.)
We thus find four solutions. Specifically,
\begin{equation}
a_{3,4}^*=\pm\left(\frac{\Psi_4}{\Psi_0}\right)^{1/4}\; \; \; \; \; 
b_{3,4}=
\mp\frac{1}{2}\left(\frac{\Psi_0}{\Psi_4}\right)^{1/4}
\end{equation}
\begin{equation}
a_{5,6}^*=\pm
i\left(\frac{\Psi_4}{\Psi_0}\right)^{1/4}\; \; \; \; \; b_{5,6}=
\pm \frac{i}{2}\left(\frac{\Psi_0}{\Psi_4}\right)^{1/4}\, .
\end{equation}
The corresponding null vectors are 
\begin{equation}
\ell_{3,4}''=\frac{1}{4}\ell\mp\frac{1}{4}\frac{{\Psi_0^*}^{1/4}}{{\Psi_4^*}^{1/4}}m
\mp\frac{1}{4}\frac{{\Psi_0}^{1/4}}{{\Psi_4}^{1/4}}{\bar m}+\frac{1}{4}
\frac{{(\Psi_0\Psi_0^*)}^{1/4}}{{(\Psi_4\Psi_4^*)}^{1/4}}n
\end{equation}
\begin{equation}
n_{3,4}''=n\pm\frac{{\Psi_4}^{1/4}}{{\Psi_0}^{1/4}}m\pm
\frac{{\Psi_4^*}^{1/4}}{{\Psi_0^*}^{1/4}}{\bar m}+
\frac{{(\Psi_4\Psi_4^*)}^{1/4}}{{(\Psi_0\Psi_0^*)}^{1/4}}\ell
\end{equation}
\begin{equation}
\ell_{5,6}''=\frac{1}{4}\ell\mp\frac{i}{4}\frac{{\Psi_0^*}^{1/4}}{{\Psi_4^*}^{1/4}}m
\pm\frac{i}{4}\frac{{\Psi_0}^{1/4}}{{\Psi_4}^{1/4}}{\bar m}+\frac{1}{4}
\frac{{(\Psi_0\Psi_0^*)}^{1/4}}{{(\Psi_4\Psi_4^*)}^{1/4}}n
\end{equation}
and
\begin{equation}
n_{5,6}''=n\pm i\frac{{\Psi_4}^{1/2}}{{\Psi_0}^{1/4}}m\mp i
\frac{{\Psi_4^*}^{1/4}}{{\Psi_0^*}^{1/4}}{\bar m}+
\frac{{(\Psi_4\Psi_4^*)}^{1/4}}{{(\Psi_0\Psi_0^*)}^{1/4}}\ell\, .
\end{equation}
    
Again, we find that 
$n_3''=K_2\ell_4''$, $\ell_3''=K_2^{-1}n_4''$,
$n_5''=K_3\ell_6''$ and $\ell_5''=K_3^{-1}n_6''$, where 
$K_2=K_3=4(\Psi_4\Psi_4^*)^{1/4}/(\Psi_0\Psi_0^*)^{1/4}.$ That is,
the four frames are just {\em two} additional distinct frames, where we
interchange the roles of $\ell,n$ (up to a scale factor).
The Weyl scalars in the new frame are
\begin{eqnarray}
\Psi_0''&=& 
\left[\Psi_0\Psi_4^3+6{a^*}^2\Psi_4\Psi_2(9\Psi_2^2+\Psi_4\Psi_0)+
{a^*}^4(81\Psi_2^4\right. \nonumber \\
&-&\Psi_4^2\Psi_0^2
+54\Psi_0\Psi_4\Psi_2^2)+6{a^*}^6\Psi_0
\Psi_2(9\Psi_2^2+\Psi_4\Psi_0) \nonumber
\\
&+&\left. {a^*}^8\Psi_4\Psi_0^3\right] /
(\Psi_4+6{a^*}^2\Psi_2+{a^*}^4\Psi_0)^3\, ,
\end{eqnarray}
\begin{equation}
\Psi_2''=\frac{\Psi_4\Psi_2-3{a^*}^2\Psi_2^2+{a^*}^4\Psi_2\Psi_0+
{a^*}^2\Psi_0\Psi_4}{\Psi_4+6{a^*}^2\Psi_2+{a^*}^4\Psi_0}\, ,
\end{equation}
and
\begin{equation}
\Psi_4''=\Psi_4+6{a^*}^2\Psi_2+{a^*}^4\Psi_0\, .
\end{equation}

Note, that $\Psi_0''$, $\Psi_2''$ and $\Psi_4''$ are unchanged if we
choose $a^*_3$, $a^*_5$ or $a^*_4$, $a^*_6$, respectively, to get the two
new frames, because $\Psi_0''$,
$\Psi_2''$ and $\Psi_4''$ are even functions of $a^*$. In particular, the
product $\Psi_0''\Psi_4''$ is invariant under the change of sign in
$a^*$. On the other hand, if we change $a^*$ by a multiplication by $i$,
i.e., change $a^*_3$ to $a^*_5$ (or $a^*_4$ to $a^*_6$) the Weyl scalars
will in general change, because they include terms which are not quartic
in $a^*$. 

We showed that we can find all the three distinct transverse null frames
for type-I space-times, and in general the product $\Psi_0''\Psi_4''$ will
be different in these three transverse frames. The above analysis allows
us to find all the three unique radiation scalars $\Psi_0\Psi_4$ in all
the transverse frames of type-I space-times.

\section{Transverse frames for algebraically special space-times}
\label{app:as}

Algebraically special space-times are not likely to arise in numerical
simulation, unless sought explicitly. For completeness, we discuss in
this Appendix transverse frames in algebraically special space-times.

\subsection{Type-II}
We can always find a standard form frame in which only $\Psi_2$ and
$\Psi_3$
are non-zero. In that frame do a class I null rotation with parameter $a$
and a subsequent class II rotation with parameter $b$. Demanding that in
the new frame $\Psi_3''=0$ implies that
\begin{equation}
b=-\frac{1}{2a^*}\frac{\Psi_3+3a^*\Psi_2}{2\Psi_3+3a^*\Psi_2}\, .
\end{equation}
Then, $\Psi_1''=0$ if either $\Psi_3=0$ (type-D), or if 
\begin{equation}
a^*=-\frac{\Psi_3}{2\Psi_2}\, .
\end{equation}
Thus we find that there is a {\em unique} transverse frame (up to
rotations in the $m,{\bar m}$ plane). In that frame, $\Psi_0''=0$,
$\Psi_1''=0$, $\Psi_2''=\Psi_2$, $\Psi_3''=0$, and
$\Psi_4''=-\frac{2}{3} \Psi_3^2/\Psi_2$, such that $\Psi_0''\Psi_4''=0$.

\begin{table}
\caption{The number of distinct equivalency classes of transverse frames
in space-times of each Petrov type and the Weyl scalars in the transverse
frames in terms of the Weyl scalars of the standard forms (see text). For
type-I space-times we list the Weyl scalars in Appendix \ref{app:mc}. In
the case of Petrov type-D we emphasize that we have the singled out
Kinnersley frame in addition to infinitely many non-Kinnersley
frames. (The table lists the Weyl scalar only for the non-Kinnersley
cases.) In this Table the unprimed Weyl scalars are in the standard form 
frames, and the double-primed scalars are in the transverse frames
(TFs).} \begin{ruledtabular} \begin{tabular}{ccccccc}
Petrov    & No. of & $\Psi_0''$ & $\Psi_1''$ &$\Psi_2''$ &$\Psi_3''$
&$\Psi_4''$  
\\
type & TFs & & & & & \\
\hline
I & $3$ &  & 0 &  & 0 &  \\
D    & $\infty\oplus 1$ & $\frac{3}{8}\Psi_2/{a^*}^2$ & 0&
$-\frac{1}{2}\Psi_2$ &
 0 & $6{a^*}^2\Psi_2$ \\
II    & $1$ & 0 & 0 & $\Psi_2$ & 0 & $-\frac{2}{3} \Psi_3^2/\Psi_2$ \\
III & $0$ &-- &-- &-- &-- &--  \\
N & $\infty$ & $0$ & 0& 0& 0& $\Psi_4$  \\
0       & $\infty$  & $0$ & 0& 0& 0& 0  
\end{tabular}
\end{ruledtabular}
\end{table}

\subsection{Type-D}
We can always find a standard form frame in which only $\Psi_2$ is
non-zero. Notice, that this is already a transverse frame. In fact, this
is the Kinnersley frame, in which the real null vectors coincide with
the directions of the (repeated) principal null directions of the
Weyl tensor. For any non-zero $a$, if we
choose $b=-1/(2a^*)$, both the new $\Psi_1''$ and $\Psi_3''$ will be
zero. That is, there is an {\em infinite number} of transverse frames. We
can parametrize all these frames with $a^*$. In all these frames
$\Psi_0''=\frac{3}{8}\Psi_2/{a^*}^2$, $\Psi_1''=0$,
$\Psi_2''=-\frac{1}{2}\Psi_2$, $\Psi_3''=0$, and
$\Psi_4''=6{a^*}^2\Psi_2$, such that in all these frames the product
$\Psi_0''\Psi_4''=\frac{3}{2}\Psi_2^2$ is independent of $a^*$.
Notice that among the infinitely many transverse frames for type-D
space-times, there is a unique frame that is singled out, specifically,
the Kinnersley frame. In the Kinnersley frame the radiation scalar
vanishes, whereas in the continuum of non-Kinnersley transverse frames
the radiation scalar is non-zero.

\subsection{Type-III}
We can always find a standard form frame in which only $\Psi_3$ is
non-zero. If we
choose $b=-1/(4a^*)$ we can make $\Psi_3''=0$, but then $\Psi_1''\ne 0$
(unless $\Psi_3=0$, which is type-0). Alternatively, we can choose
$b=-3/(4a^*)$ which makes $\Psi_1''=0$, but  $\Psi_3''\ne 0$ (unless it is
type-0). That is, we cannot nullify both $\Psi_1''=0$ and $\Psi_3''=0$
simultaneously. There are {\em no} transverse frames for type-III
space-times.

\subsection{Type-N}
We can always find a standard form frame in which only $\Psi_4$ is
non-zero. Note, that
this is already a transverse frame.  No matter which $a^*$ we choose, we
remain in a transverse frame. That is, there is an {\em infinite number} 
of transverse frames, in all of which $\Psi_0''\Psi_4''=0$.

\subsection{Type-0}
In type-0 space-times all the Weyl scalar are zero, and all null rotations
will preserve this. There are {\em infinitely many} transverse null
frames.

\end{appendix}

%%%%%%%%%%%%%%%%%%%%%%
%%%   REFERENCES   %%%
%%%%%%%%%%%%%%%%%%%%%%

\bibliographystyle{apsrev}
\bibliography{AN10}

%%%%%%%%%%%%%%%
%%%   END   %%%
%%%%%%%%%%%%%%%

\end{document}